\documentclass[
reprint, showpacs, bibnotes, amsmath, amssymb, aps, pra
]{revtex4-1}

\usepackage{graphicx}
\usepackage{dcolumn} 
\usepackage{bm}
\usepackage{xspace}
\usepackage{amsfonts}
\usepackage{siunitx} 
\usepackage{xcolor}
\usepackage{tikz}
\tikzset{base/.style = {rectangle,  align=center,
 inner xsep=0ex, inner ysep=0ex},
 whitebox/.style = {base, fill=white, inner sep=0ex},
 }
\usetikzlibrary{calc}
\usepackage[colorlinks=True,linkcolor=blue,citecolor=blue]{hyperref}

\newcommand{\Tc}{\ensuremath{T_\text{c}}\xspace}
\newcommand{\Rm}{\ensuremath{R_\text{m}}\xspace}
\newcommand{\kB}{\ensuremath{k_\text{B}}\xspace}
\newcommand{\Gh}{\ensuremath{\mathit{\Gamma}_\text{h}}\xspace}

\begin{document}

\preprint{APS/123-QED}
\title{Observation of superconductivity and surface noise\\using a single trapped ion as a field probe}

\author{K. Lakhmanskiy$^{1,\star}$, P. C. Holz$^{1,\star}$, D. Sch\"artl$^1$, B. Ames$^1$, R. Assouly$^1$,\\T. Monz$^1$, Y. Colombe$^1$, R. Blatt$^{1,2}$}
\affiliation{%
\mbox{$^1$Institut f\"ur Experimentalphysik, Universit\"at Innsbruck, Technikerstrasse 25, 6020 Innsbruck, Austria}
\mbox{$^2$Institut f\"ur Quantenoptik und Quanteninformation,}
\mbox{\"Osterreichische Akademie der Wissenschaften, Technikerstr. 21\,A, 6020 Innsbruck, Austria}
$^\star$these authors contributed equally to this work
}

\date{\today}

\begin{abstract}

Measuring and understanding electric field noise from bulk material and surfaces is important for many areas of physics. In this work, we demonstrate the probing of electric field noise from different sources with an ion, $\SI{225}{\micro\meter}$ above the trap surface. We detect noise levels as small as $S_E = \SI{5.2 \pm 1.1 e-16}{\volt^2\meter^{-2}\hertz^{-1}}$ at $\omega_{\text{z}} =2\pi\times\SI{1.51}{\mega\hertz}$ and $T=\SI{12}{\kelvin}$, the lowest noise level observed with a trapped ion to our knowledge. Our setup incorporates a controllable noise source utilizing a high-temperature superconductor. This element allows us, first, to benchmark and validate the sensitivity of our probe. Second, to probe non-invasively bulk properties of the superconductor, observing for the first time a superconducting transition with an ion. For temperatures below the transition, we use our setup to assess different surface noise processes. The measured surface noise shows a deviation from a power-law in the frequency domain. However, the temperature scaling of the data is not in a good agreement with existing surface noise models. Our results open perspectives for new models in surface science and pave the way to test them experimentally.

\end{abstract}

\maketitle

\section{Introduction}
Electric field noise provides insights into microscopic processes, and imposes limitations to experimental systems. In particular, electric field noise in close proximity to surfaces creates obstacles for near-field measurements \cite{EFN1,CCD}, experiments with nitrogen-vacancy centers \cite{Kim15}, Casimir effect studies \cite{EFN2}, gravitational-wave detectors \cite{EFN3}, and ion trapping experiments \cite{Bro15}. It has been suggested to employ the high sensitivity of trapped ions to electric field noise as a new tool in surface science \cite{Hit13}. Trapped ions have been used to study the dependence of electric field noise on frequency, trap temperature and ion-surface distance \cite{Tur00,Des06,Lab08,Chi15,Sed17,Bol17,Hit17} and have been combined with the analysis and removal of surface contaminants \cite{Hit12,Dan14,Hit17}. In this work, we use a surface-electrode ion trap containing a high-temperature superconductor to investigate not only surface noise but also bulk material properties. We operate the trap in two distinct regimes, above and below the critical temperature \Tc of its superconducting electrodes. Above \Tc, the electric field noise sensed by the ion originates from the bulk resistance of two long electrodes; below \Tc, this resistance vanishes and the ion probes the noise from the surface of the trap. In this way, we compare different sources of electric field noise in situ, with a single device. The capability to probe the resistivity of the superconductor non-invasively with an ion also allows us to observe the superconducting transition without direct electrical probing. This constitutes the first observation of superconductivity using an ion as a probe. Conventional superconductors have been used in the past as ion trap material to study electric field noise above and below \Tc \cite{Shan2010,Chi2014}. In these studies, however, the onset of superconductivity did not lead to a measurable modification of the electric field noise at the ion.

Important sources of electric field noise in trapped ion experiments are technical noise, Johnson-Nyquist (Johnson) noise, and surface noise. Technical noise is related to control devices like power supplies as well as to electromagnetic interference from nearby electronics. Johnson noise is caused by thermal motion of charge carriers in conductors \cite{Joh28}. Surface noise is thought to arise from different physical processes related to the surface material \cite{Bro15}. We measure the frequency spectrum and temperature dependence of the electric field noise to differentiate between these noise sources.

\section{Setup}
\begin{figure}[htbp]
\centering
\includegraphics[width=0.4\textwidth]{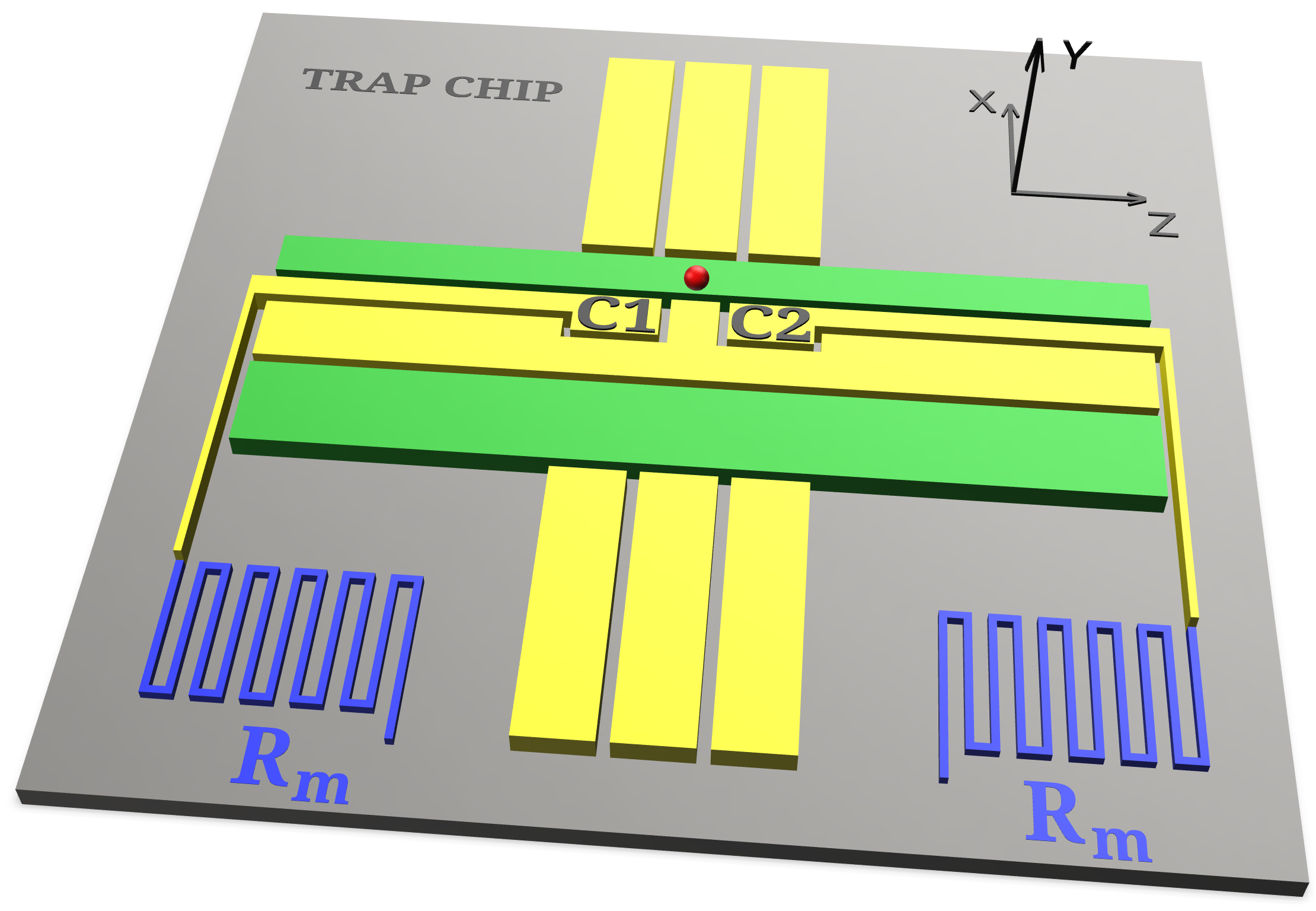}
\caption{Schematic illustration of the trap electrodes. DC (yellow) and RF (green) electrodes confine a single ion (red sphere) in the trap center above the surface. The trap electrodes are made out of YBCO and covered with gold. Two central DC electrodes C1, C2 are connected to meander resistors \Rm (blue), made only of YBCO and integrated to the trap chip.}
\label{fig:trap-schematic}
\end{figure}

Our single ion probe is confined in a linear surface-electrode Paul trap (Fig.\,\ref{fig:trap-schematic}).
A sapphire substrate supports \SI{50}{\nano\meter}-thick electrodes made of $\text{YBa}_2\text{Cu}_3\text{O}_7\,$ (YBCO), a high-temperature superconductor with a critical temperature $\Tc \approx \SI{85}{\kelvin}$. To ensure operability of the trap above \Tc the YBCO electrodes are covered with \SI{200}{\nano\meter} of gold. The key feature of the trap is a pair of electrodes C1 and C2 near the trap center, connected to two identical meander-shaped structures. These meanders are made of YBCO only, without gold coating. Below \Tc the resistance \Rm of each meander is negligible. Above \Tc the meanders' resistance \Rm gives rise to Johnson noise, which translates to electric field noise at the trap center that can be sensed with an ion. This noise source can be switched on and off by adjusting the trap chip temperature. The geometry of electrodes C1 and C2 is designed such that electric fields from correlated voltages cancel out at the center of the trap, $\bm{E}^\text{(C1)} (\bm{r}=0) = -\bm{E}^\text{(C2)}(\bm{r}=0)$, which minimizes the influence of pickup from the RF electrode by C1 and C2. However, the uncorrelated Johnson noise in the meanders adds up, leading to an electric field noise $S_E = S_E^\text{(C1)} + S_E^\text{(C2)}$.

The trap chip is mounted on a heatable copper stage that is thermally isolated from the environment. The trap chip temperature, measured with a Si diode sensor, can be set in the range $T = (10 - 200)\,$K, while the low-pass filter boards and RF resonator stay at a nearly constant temperature $T_{\text f} \approx (10 - 14)\,$K. This thermal decoupling ensures that noise from off-chip sources, e.\,g., Johnson noise from the low-pass filters or external technical noise attenuated by the filters, is nearly independent of the trap chip temperature. We determine the critical temperature \Tc by means of a 4-wire measurement of \Rm using a third on-chip YBCO meander (not shown in Fig.\,\ref{fig:trap-schematic}) identical to the ones connected to C1 and C2. This DC measurement of \Rm is used to calculate the Johnson noise in the MHz regime for $T > \Tc$ where the skin depth $\zeta$ is orders of magnitude larger than the YBCO film thickness (appendix \ref{app:skin-depth}).

The experiment is performed in a cryogenic apparatus \cite{Nie14,Nie15}. We confine a single $^{40}\text{Ca}^+$ ion at a distance $d =\SI{225}{\micro\meter}$ above the surface of the trap chip using static (DC) and radio-frequency (RF) electric fields. An RF drive voltage $V_{\text{RF}} \sim \SI{230}{\volt} $ at $\omega_{\text{RF}} =2\pi\times\SI{17.6}{\mega\hertz}$ provides radial confinement $\omega_{x,y}\sim 2\pi\times\SI{3}{\mega\hertz}$ in the $xy$ plane. The axial motional frequency $\omega_z$ is varied in the range $\omega_z = 2\pi\times (0.4-1.8)\,\si{\mega\hertz}$ by changing the DC voltages. 
Electric field noise couples to the ion and adds phonons to its motional state at a rate \Gh. The relation between this heating rate \Gh and the electric field noise spectral density $S_E(\omega)$ at the position of the ion is \cite{Bro15}
\begin{equation}
\label{eq:HR-definition}
\Gh=\frac{q^2}{4m\hbar\omega}S_E(\omega)\,,
\end{equation}
with $\hbar$ the reduced Planck constant, $q$ and $m$ the ion's charge and mass, and $\omega$ its motional frequency. The ion is prepared in the ground state of its axial mode by Doppler and subsequent sideband laser cooling. A narrow linewidth \SI{729}{\nano\meter} laser tuned to the $S_{{1}/{2}} \leftrightarrow D_{{5}/{2}}$ quadrupole transition is used to measure \Gh with the sideband ratio method \cite{Leib03}. The measurement uncertainties of \Gh in our experiments are limited by quantum projection noise \cite{QPN}.

\section{Results}

\begin{figure}[htbp]
	\centering
	\begin{tikzpicture}
		\node[anchor=south west, inner sep=0] (X) at (0\textwidth,0){\includegraphics[width=0.48\textwidth]{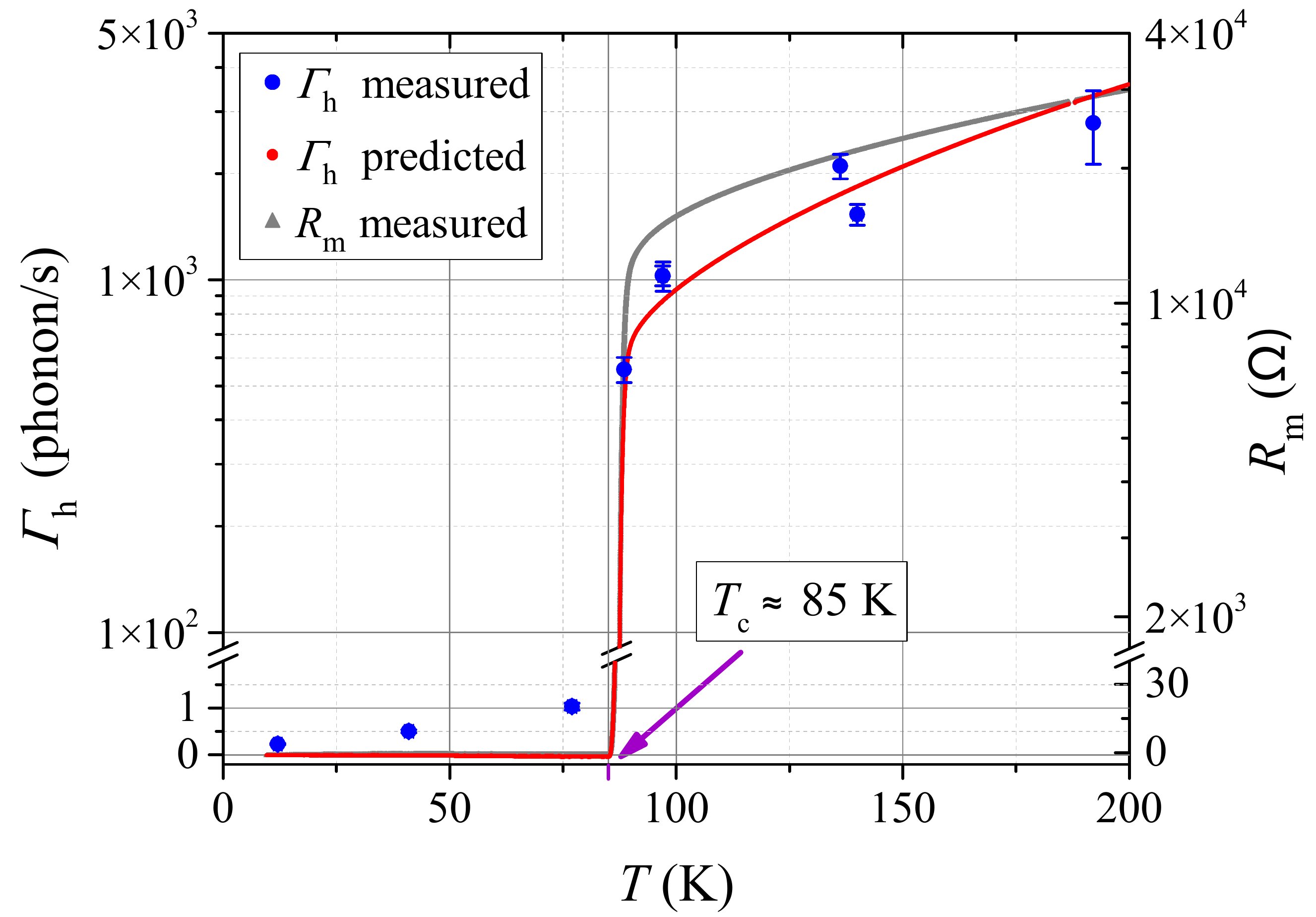}};       
  	\end{tikzpicture}
  	\caption{Observation of the superconducting transition of YBCO with a trapped ion. Blue dots show the measured ion motional heating rate \Gh as a function of trap chip temperature $T$ for a trap frequency $\omega_z \approx 2\pi \times\SI{1.0}{\mega\hertz}$. The measured meander resistance \Rm (gray data) is used to calculate the motional heating rate expected from Johnson noise in the meanders connected to C1 and C2 (red data). Note the break in the vertical axes.}
  \label{fig:HR-vs-temp}
\end{figure}

In a first study, we detect non-invasively the superconducting transition of YBCO using a single trapped ion as a probe. For this, the ion's heating rate \Gh is measured for different trap chip temperatures while keeping the axial frequency constant $\omega_z \approx 2\pi \times\SI{1.0}{\mega\hertz}$, see Fig.\,\ref{fig:HR-vs-temp}.
Below \Tc, the heating rate increases slowly from $\Gh = \SI{0.23 \pm 0.02}{phonons\per\second}$ to $\Gh = \SI{1.03 \pm 0.08}{phonons\per\second}$ between $T = \SI{12}{\kelvin}$ and $T = \SI{77}{\kelvin}$. From $T = \SI{77}{\kelvin}$ to $T = \SI{89}{\kelvin}$ the heating rate increases by roughly a factor 500 to $\Gh = \SI{556 \pm 46}{phonons\per\second}$. This sudden increase coincides with the superconducting transition at $\Tc\approx\SI{85}{\kelvin}$, as evidenced by the 4-wire resistance measurement (Fig.\,\ref{fig:HR-vs-temp}, gray data). For $T > \Tc$, we show that the ion heating rate corresponds to what is expected from Johnson noise in the YBCO meanders connected to C1 and C2. The electric field spectral density of Johnson noise is given by \cite{Joh28,Nyq28,Bro15}
\begin{equation}
\label{eq:JN}
S_E^\text{(JN)} = \frac{4\kB T R(\omega,T)}{\delta_\text{c}^2}\,,
\end{equation}
where \kB is Boltzmann's constant, $T$ the temperature of the resistor causing the noise, $R$ its resistance, and $\delta_\text{c}$ a geometry-dependent characteristic distance \cite{Bro15}. We calculate $\delta_\text{c}=\SI{5.1}{\milli\meter}$ for electrodes C1 and C2 from trap simulations \cite{trap_sim}. Since the meanders are located directly on the trap chip, filter effects can be neglected, i.e., $R(\omega,T) = \Rm(T)$. Based on the resistance and temperature measurements we calculate the expected heating rate from Eqs.\,(\ref{eq:HR-definition}, \ref{eq:JN}) (Fig.\,\ref{fig:HR-vs-temp}, red data). The measured heating rates are in good agreement with the expected values, with an average deviation $\overline{\Delta} = 1.9$. $\overline{\Delta} = \langle |\mathit{\Gamma}_\text{h}^\text{(meas)} - \mathit{\Gamma}_\text{h}^\text{(exp)}|/\sigma\rangle\,$, where $\mathit{\Gamma}_\text{h}^\text{(meas)}$ and $\mathit{\Gamma}_\text{h}^\text{(exp)}$ are the measured and expected heating rates, and $\sigma$ is the standard deviation of an individual data point.\\

\begin{figure}[htbp]
	\centering
	\begin{tikzpicture}
		\node[anchor=south west, inner sep=0] (X) at (0\textwidth,0){\includegraphics[width=0.46\textwidth]{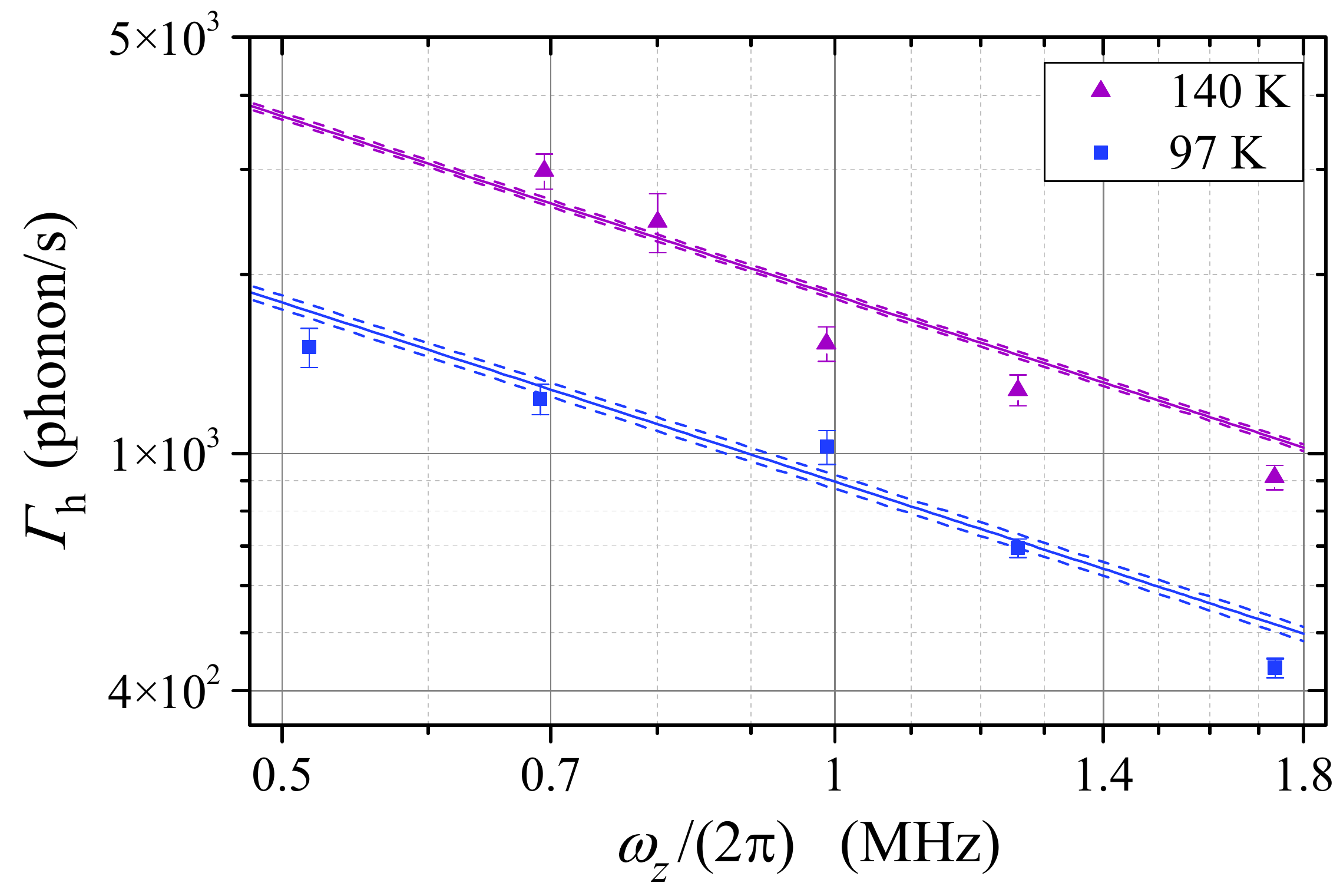}};      
  	\end{tikzpicture}
  	\caption{Characterization of the on-chip white noise source above \Tc using a trapped ion. Blue and purple dots show measured heating rate \Gh as function of trap frequency for trap chip temperatures $T = (97,\,140)\,\text{K} > \Tc$. Solid lines are predictions for Johnson noise from the meander resistance \Rm. Dashed lines reflect the \SI{1}{\kelvin} uncertainty in the temperature measurement.}
  \label{fig:HR-vs-freq-above}
\end{figure}

In a second study, we measure the spectrum of the electric field noise for trap chip temperatures above and below \Tc. Above the transition we confirm the white noise nature of our temperature-switchable on-chip noise source. For this, the heating rate is measured as a function of the trap frequency $\omega_z$ for two different temperatures $T = \SI{97}{\kelvin}$ and $T = \SI{140}{\kelvin}$, see Fig.\,\ref{fig:HR-vs-freq-above}.
The solid lines show the predicted heating rate calculated from the measured resistance \Rm using equations \eqref{eq:HR-definition},\eqref{eq:JN}. The measured data show good agreement with the calculated curves with an average deviation $\overline{\Delta} = 2.06$ for $T = \SI{97}{\kelvin}$ and $\overline{\Delta} = 2.12$ for $T = \SI{140}{\kelvin}$. This validates the sensitivity of our probe. We note that there exists another way to certify the sensitivity which uses noise injection to one of the trap electrodes \cite{Sch15,Bla09,Sed18,Dom18}. Our method has the advantage that the white noise source is placed directly on chip and is therefore unfiltered.

\begin{figure}[htbp]
\centering
\includegraphics[width=0.43\textwidth]{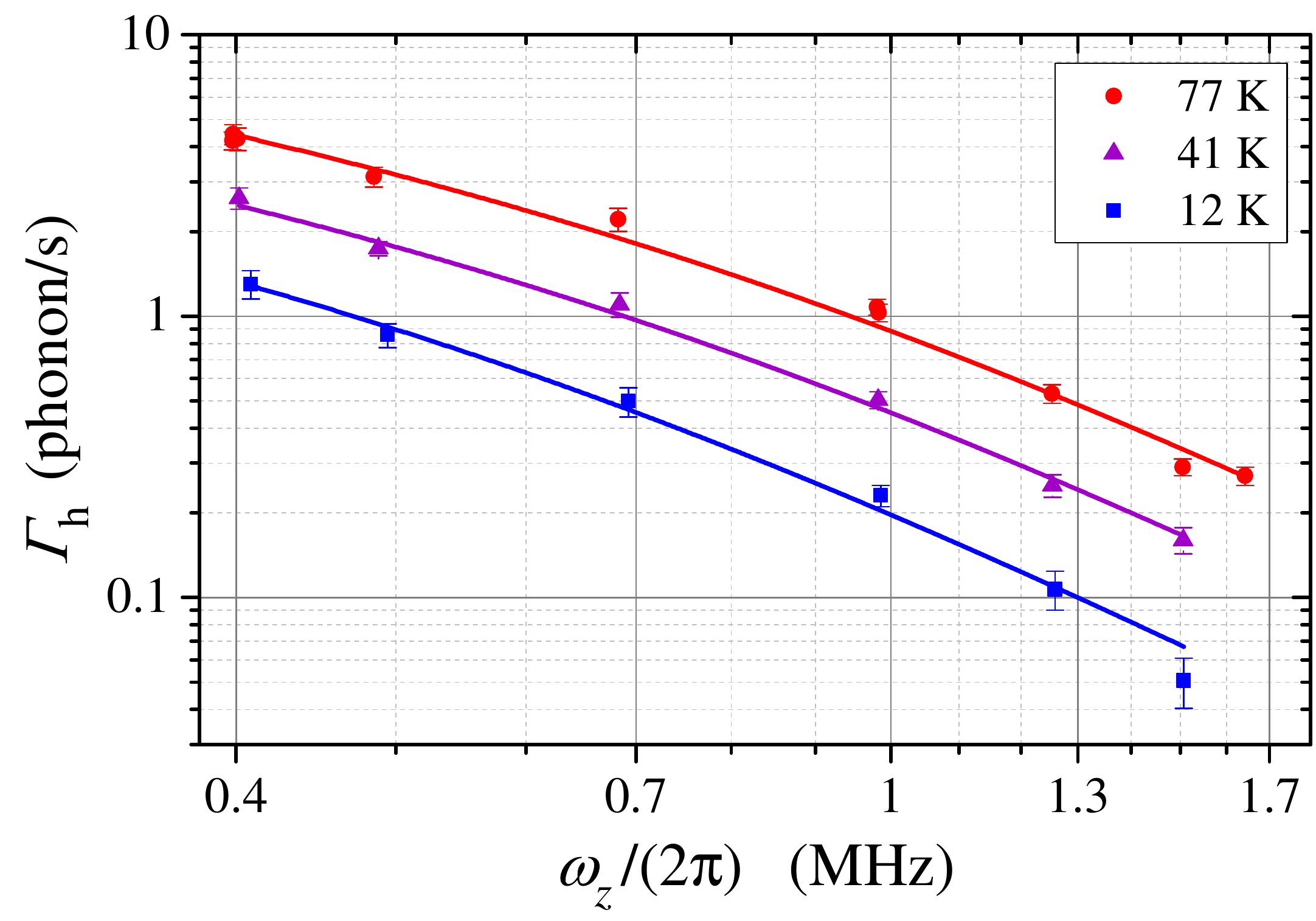}
\caption{Characterization of the surface noise below \Tc using a trapped ion. Blue, purple and red dots show measured heating rate as function of trap frequency for trap chip temperatures $T = (12,\,41,\,77)\,\text{K} < \Tc$. The solid lines are a fit to the data with the TLF model, Eq.\,\eqref{eq:TLF}.}
\label{fig:HR-vs-freq-below}
\end{figure}

For $T < \Tc$, the heating rate spectrum is measured at three different temperatures $T = (12,\,41,\,77)\,$K (Fig.\,\ref{fig:HR-vs-freq-below})\,\footnote{datasets at temperatures $T = (12,\,41)\,$K were each taken over the course of one day, the dataset at temperature $T = \SI{77}{\kelvin}$ was taken over the course of 3 days}. The lowest measured heating rate is $\Gh = \SI{0.051 \pm 0.010}{phonons\per\second}$ at $T=\SI{12}{\kelvin}$ and $\omega_z = 2\pi\times\SI{1.51}{\mega\hertz}$ which corresponds to an electric field spectral density $S_E = \SI{5.2 \pm 1.1 e-16}{\volt^2\meter^{-2}\hertz^{-1}}$, see Eq.\,\eqref{eq:HR-definition}. To our knowledge this is the lowest electric field noise measured with a trapped ion to date \cite{Bro15}. 

To confirm that the main origin of the measured ion heating rate for $T < \Tc$ is surface noise, we exclude other possible noise sources. Specifically, we rule out external technical noise which is independent of the trap chip temperature, in contrast to the measured heating rates. Johnson noise from filters, wiring and trap electrodes is calculated to be significantly smaller than the noise we measure. Finally, repeating the experiment without superconducting YBCO meanders shows that these do not contribute to the heating rate for $T < \Tc$ within the uncertainty of our measurement. The above arguments, detailed in appendices \ref{app:technical-noise}, \ref{app:Johnson-noise} and \ref{app:meanders}, imply that the noise causing the ion's motional heating below \Tc originates at the surface of the trap.\\

In the remainder of this section we analyze the spectral properties and temperature dependence of the surface noise observed below \Tc (Fig.\,\ref{fig:HR-vs-freq-below}). As main result we will show that the measured surface noise spectrum deviates from a power law. To do this we fit the data both with a power law and with a two-level fluctuator (TLF) model, that predicts a crossover dependence in frequency. The power law is given by 
\begin{equation}
\label{eq:HR-power-law}
\Gh= c\, \omega_z^{-\alpha}\,.
\end{equation}
We find a power-law exponent $\alpha\approx 2$ for all three data sets, corresponding to a $1/f$ frequency scaling of the electric field noise $S_E$. The exponent is close to the ones reported in Refs.\,\cite{Sed17,Bol17}, where a $1/d^4$ distance scaling of the heating rate was found, indicative of surface noise. However, a detailed analysis of the frequency dependence in the data of Fig.\,\ref{fig:HR-vs-freq-below} reveals a change in the local power law exponent $\alpha$ around $\SI{0.8}{\mega\hertz}$, which indicates a crossover between low- and high-frequency domains. This behavior is predicted by TLF models. TLF models consider real or effective particles undergoing random transitions between two quantum states with different electric dipole moment. Transitions between the TLF states at a rate $\omega_0$ induced either by thermal activation or quantum tunneling lead to electric field fluctuations with a spectral density \cite{Bro15}
\begin{equation}
\label{eq:TLF}
S_E^\text{(TLF)}(\omega) = A\frac{\omega_0}{\omega_0^2+\omega^2} \,.
\end{equation}
The spectrum, eq.\,\eqref{eq:TLF}, also approximately describes the noise from fluctuating dipoles of adatoms adsorbed to the trap surface \cite{Bro15}. The solid lines in Fig.\,\ref{fig:HR-vs-freq-below} show a TLF fit to the heating rate data below \Tc. The TLF model shows a significantly better agreement with our data than the power law for all three temperature sets, as evidenced by the reduced chi squared values in Tab.\,\ref{tab:power-law}. Two adjustable parameters are used for both models. This proves that the measured noise spectrum deviates from a power-law.
\begin{table}[ht]
\caption{Statistical evidence of the deviation from a power law of the measured heating rate spectra below \Tc, Fig.\,\ref{fig:HR-vs-freq-below}. The second column shows the reduced chi squared for the power law fit, the third column shows the reduced chi squared for the TLF model fit.}
\label{tab:power-law}
\begin{ruledtabular}
\begin{tabular}{ccccc}
 & $T ~ (\SI{}{\kelvin})$ & $\chi^2_\text{power law}$ & $\chi^2_\text{TLF model}$ & \\ 
\colrule
 & 12  & 2.9 & 1.3 & \\
 & 41  & 2.4 & 0.8 & \\
 & 77  & 6.0 & 2.3 & \\
\end{tabular}
\end{ruledtabular}
\end{table}

The TLF fit parameters are presented in Tab.\,\ref{tab:TLF fit parameters}. We find the crossover frequency in the range $\omega_0 = 2\pi\times (0.6-0.8)\,\si{\mega\hertz}$, with a slight dependence on the temperature. The dominant temperature dependence of the spectrum given by Eq.\,\eqref{eq:TLF}, scales as \cite{Bro15}
\begin{equation}
\label{eq:TLF-mag}
A(T)=A_0\cosh^{-2}(T_0/2T) \,.
\end{equation}
This dependence cannot be matched with the measured temperature scaling of the spectra in Fig.\,\ref{fig:HR-vs-freq-below} \footnote{The adatom dipole fluctuator model predicts a temperature scaling $A(T)\approx \text{const}\cdot T^{2.5}$ when including higher excited vibrational levels \cite{Bro15}. Our data do not agree with this scaling either.}. Averaging over a distribution of fluctuators can lead to a significantly different temperature scaling \cite{Bro15}. While typical averaging procedures do not retain the crossover in the frequency dependence \cite{Bro15}, this approach might still lead to a model that is consistent with our data.
\begin{table}[ht]
\caption{Crossover frequency $\omega_0(T)$ and magnitude prefactor $A(T)$ resulting from the TLF model fit, Eq.\,\eqref{eq:TLF}, to the spectral data in Fig.\,\ref{fig:HR-vs-freq-below}.}
\label{tab:TLF fit parameters}
\begin{ruledtabular}
\begin{tabular}{ccccc}
 & $T ~ (\SI{}{\kelvin})$ & $\omega_0/(2\pi) ~ (\SI{}{\mega\hertz})$ & $A \times 10^8 ~ (\SI{}{\volt^2\meter^{-2}})$ & \\ 
\colrule
 & 12  & \SI{0.58 \pm 0.08}&\SI{2 \pm 0.1 } & \\
 & 41  & \SI{0.74 \pm 0.05}&\SI{4.1 \pm 0.1 } & \\
 & 77  & \SI{0.81 \pm 0.08}&\SI{7.8 \pm 0.3 } & \\
\end{tabular}
\end{ruledtabular}
\end{table}

Apart from the TLF and adatom dipole fluctuator model, there is to our knowledge only one other surface noise model predicting a crossover region with local power law exponent $\alpha\approx 2$. This is the adatom diffusion (AD) model. The AD model describes electric field noise arising from the diffusion of adatoms with a static dipole moment on the chip surface. In this model, the crossover frequency occurs at $\omega_0 = D/d^2$ \cite{Bro15}. For typical values of the diffusion constant $D\sim 10^{-7}\si{\meter^2\per\second}$ \cite{Zhd91} and our surface-ion separation $d = \SI{225}{\micro\meter}$, we calculate a crossover frequency $\omega_0 \sim 2\pi\times \SI{0.3}{\hertz}$ that is 6 orders of magnitude smaller than the value $\omega_0 \approx 2\pi\times\SI{0.8}{\mega\hertz} $ we observe. Diffusion of adatoms can therefore be excluded as origin of the noise that we measure.

\section{Conclusion}
In conclusion, we have used a single trapped ion as a probe for bulk and surface properties of materials, achieving the highest sensitivity to electric field noise with a single ion reported to date. Our setup incorporates an unfiltered on-chip source of white noise. We employed our ion field probe to measure non-invasively the superconducting transition of YBCO. This technique could be used in the future for the characterization of samples that cannot be subjected to a direct resistance measurement, like delicate structures or topologies that cannot be connected. For example, studies of persistent currents in arrays of metallic loops, known to be exceptionally sensitive to their environment \cite{Ble09}, might be possible. Below the transition we measured surface noise with a crossover of the power-law exponent in the frequency domain. Such a behavior is generally expected \cite{Tal16} and predicted, e.g., by TLF or adatom dipole fluctuator models, but has not been observed experimentally before. The temperature dependence of our data, however, cannot be understood with existing models. Our results, together with other recent studies of noise scaling with ion-electrode distance \cite{Sed17,Bol17} and chemical composition of surface materials \cite{Hit12,Dan14,Hit17}, gives new input for understanding the origin of surface noise. In addition, our work paves the way for the use of high-temperature superconductors for large scale ion-based quantum processors \cite{Kie02}, where low-resistance trap electrodes will become important. 

\begin{acknowledgements}
We  thank  Muir  Kumph  and  Peter  Rabl  for  discussions, and  Philipp  Schindler  and  the  quantum  information  experiment  team  for  technical  assistance.  We  acknowledge  financial  support  by  the  Austrian  Science  Fund  (FWF)  through projects P26401 (Q-SAIL) and F4016-N23 (SFB FoQuS), by the Institut für Quanteninformation GmbH, and by the Office of the Director of National Intelligence (ODNI), Intelligence Advanced  Research  Projects  Activity  (IARPA),  through  the Army Research Office Grant No. W911NF-10-1-0284. This project has received funding from the European Union’s Horizon  2020  research  and  innovation  programme  under  Grant Agreement  No.  801285  (PIEDMONS).  All  statements  off act,  opinion  or  conclusions  contained  herein  are  those  of the authors and should not be construed as representing the official  views  or policies  of IARPA,  the ODNI, or  the U.S.Government.
\end{acknowledgements}

\appendix

\section{Skin depth in YBCO for $T>\Tc$}
\label{app:skin-depth}
The skin depth $\zeta$ in a material is given by \cite{Jackson}
\begin{equation}
\label{SM:skin-depth}
\zeta=\sqrt{\frac{2\rho}{\omega\mu}}\,,
\end{equation}
where $\rho$ is the resistivity of the material, $\mu$ its permeability and $\omega$ the frequency of the applied AC electric field. We calculate the resistivity $\rho$ of our \SI{50}{\nano\meter}-thick YBCO film from the 4-wire DC resistance measurement of the meander electrode (length $\SI{5.18}{\milli\meter}$ and width \SI{10}{\micro\meter}). Taking a resistance $\Rm \approx \SI{8}{\kilo\ohm}$ of the meander electrode above \Tc (see Fig.\,\ref{fig:HR-vs-temp}), we arrive at a resistivity $\rho\approx\SI{78E-8}{\ohm\meter}$. Assuming $\mu=\mu_0=2\pi\times10^{-7}\,\si{\henry\per\meter}$ \cite{Sal02} and $\omega=2\pi\times\SI{1.8}{\mega\hertz}$ leads to a skin depth $\zeta\approx\SI{441}{\micro\meter}$, which is much larger than the YBCO film thickness.

\section{Ruling out external technical noise}
\label{app:technical-noise}
We rule out external technical noise as the origin of the ion heating rates for chip temperatures $T < \Tc$ (Fig.\,\ref{fig:HR-vs-freq-below}). We note that the measured heating rates increase with rising trap chip temperature $T$. External technical noise sensed by the ion, on the other hand, decreases with rising $T$, as we show in this section. This rules out that the measured noise is caused by technical noise for all temperature sets, except the lowest one at \SI{12}{\kelvin}. However, the \SI{12}{\kelvin} set is very likely to be dominated by the same source as the ones at higher temperature because we observe the same characteristic crossover regime in the frequency spectrum for all three temperature sets. It would be an extraordinary coincidence if the technical noise  hypothetically limiting the heating rate at the lowest temperature had the exact same spectrum.

\begin{figure}[htbp]
   \begin{tikzpicture}
	\node[] (Z) at (0,0.15\textwidth) { \small (a) };
	\node[anchor=south west, inner sep=0] (X) at (0\textwidth,0){\includegraphics[width=0.29\textwidth]{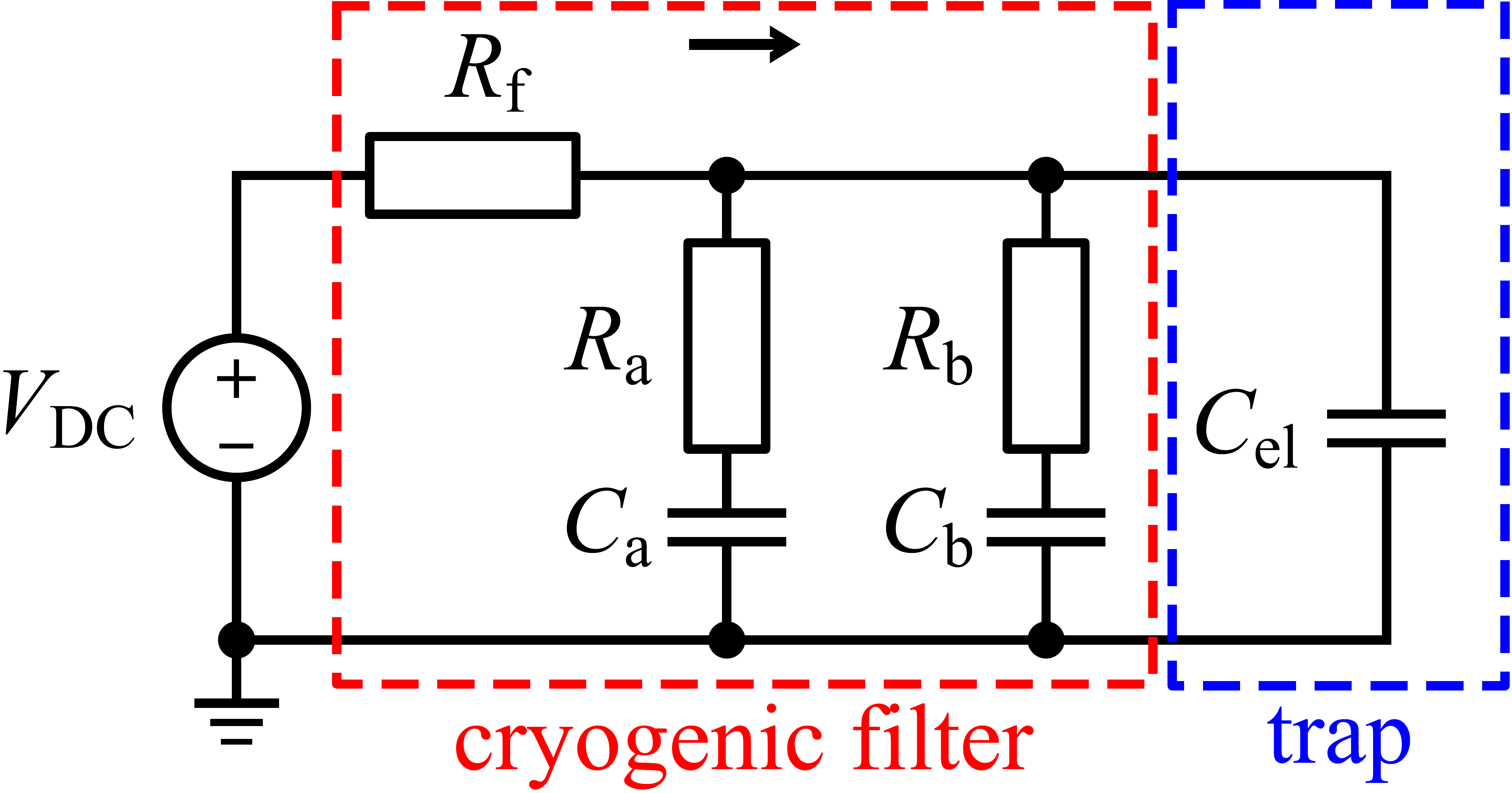}};
	\node[] (Z) at (0\textwidth,-0.02\textwidth) { \small (b) };
	\node[anchor=south west, inner sep=0] (X) at (0\textwidth,-0.17\textwidth)
{\includegraphics[width=0.29\textwidth]{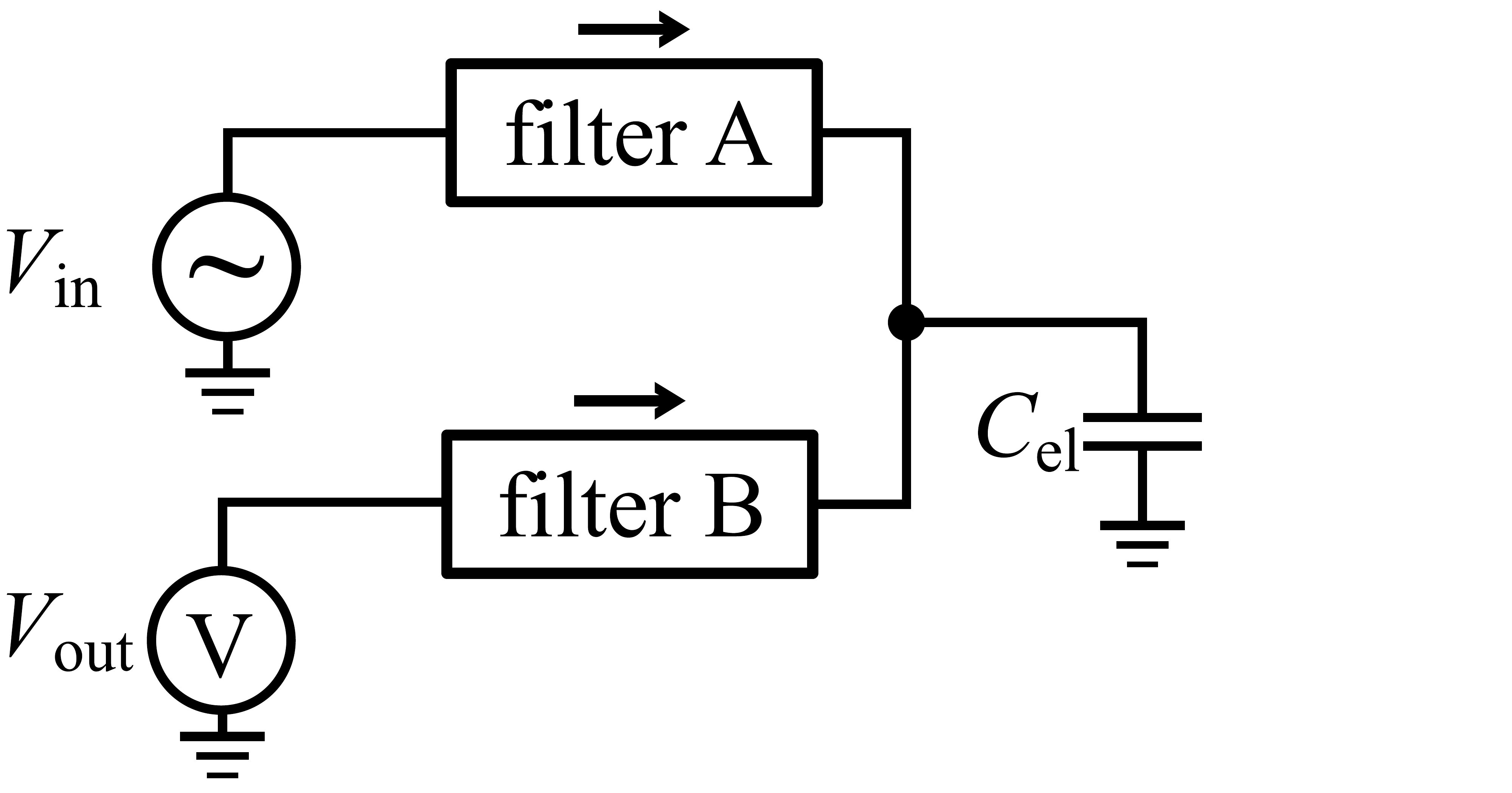}};
  \end{tikzpicture}
  \caption{(a) RC low-pass filter circuit used between the DC supplies and the trap electrodes. (b) Schematic layout of the circuit used for the measurement of the transfer function of the RC filters. The black arrows indicate the direction in which the filters act as low pass filters.}
  \label{SM:filter-layout}
\end{figure}

The thermal decoupling incorporated in our setup ensures that while we locally heat the trap chip to temperatures $T = (10 - 200)\,$K, the cryogenic environment, in particular the low-pass filters, stays at a nearly constant temperature $T_{\text f} \approx (10 - 14)\,$K. The change in $T_{\text f}$ is small, but it might still lead to a variation in the attenuation of external technical noise by the low-pass filters. Therefore, we measure the temperature dependence of the transfer function of the cryogenic low-pass filters. The filters, all identical, are placed only a few centimeters away from the trap and suppress noise that might reach the trap electrodes through the DC lines. The equivalent circuit of these first order RC filters is shown in Fig.\,\ref{SM:filter-layout}\,(a).
The filter consists of a resistor $R_\text{f}=\SI{100}{\ohm}$ (Vishay, Y1625100R000Q9R) and two capacitors $C_\text{a}=\SI{330}{\nano\farad}$ (Kemet, C2220C334J1GACTU) and $C_\text{b}=\SI{470}{\pico\farad}$ (Kemet, C0805C471J1GACTU) placed in parallel. Resistors $R_\text{a}$, $R_\text{b}$ model the equivalent series resistance (ESR) of the capacitors. The capacitance of the trap electrode to ground $C_\text{el}$ is on the order of \SI{1}{\pico\farad} and negligible compared to the filter capacitance. The electrical setup for the measurement of the filter's transfer function is shown in Fig.\,\ref{SM:filter-layout}\,(b). Two filters A and B are wire bonded to the same trap electrode. An RF signal with amplitude $V_\text{in}$ is injected into filter A, and the attenuated signal $V_\text{out}$ is measured at the input of filter B.
The transfer function measured in this configuration corresponds to that of the first order RC filter shown in Fig.\,\ref{SM:filter-layout}\,(a), however with twice the filter capacitance $C_\text{eff}\approx 2(C_\text{a}+C_\text{b})$. The additional capacitance reduces the cut-off frequency $f_\text{c}\approx\SI{4.8}{\kilo\hertz}$ by a factor of 2, which is irrelevant for the temperature scaling arguments used below. The resistance $R_\text{f}$ of filter B can be neglected due to the high input impedance of \SI{1}{\mega\ohm} of the oscilloscope used to measure the output signal $V_\text{out}$. Additional filter effects arising from $R_\text{f}$ of filter B and the outgoing cabling capacitance $C_\text{cab}\approx\SI{300}{\pico\farad}$ are negligible due to a high cut-off frequency $f_\text{c}\approx\SI{5}{\mega\hertz}$, well above the frequency range of interest.

\begin{figure}[htbp]
\centering
\includegraphics[width=0.46\textwidth]{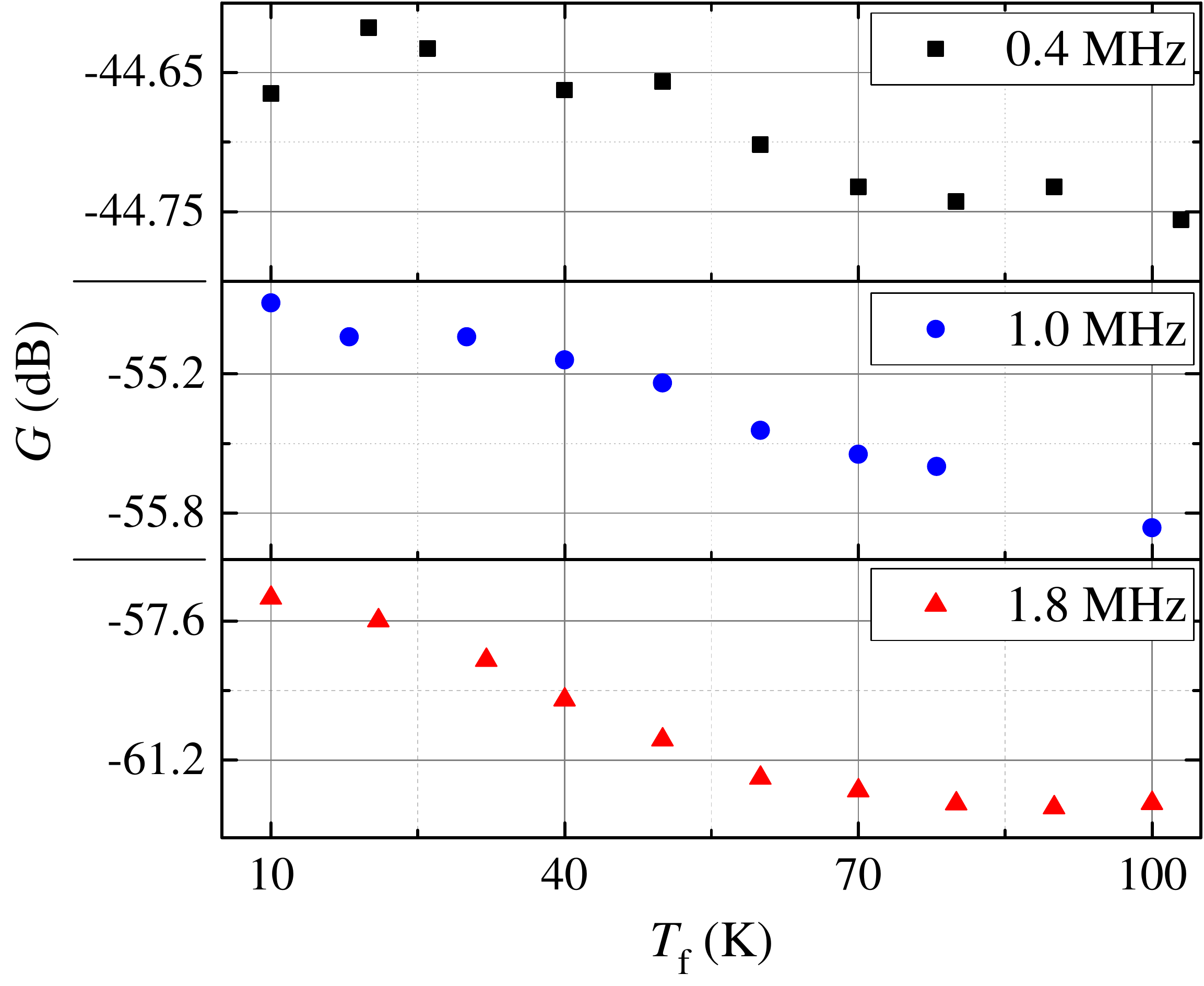}
\caption{Transfer function $G$ of the cryogenic low-pass filters measured with the setup shown in Fig\,\ref{SM:filter-layout}\,(b) as a function of the filter temperature $T_\text{f}$ at three different frequencies $\omega=2\pi\times(0.4, 1.0, 1.8)\,$MHz.}
\label{SM:filter-attenuation}
\end{figure}

Fig.\,\ref{SM:filter-attenuation} shows the filter transfer function $G=|V_\text{out}/V_\text{in}|^2$ for varying RC filter temperature $T_\text{f}$ during cooling down and warming up of the entire cryogenic apparatus. The applied change in $T_\text{f}$ strongly overestimates the variation in filter temperature  $T_{\text f} \approx (10 - 14)\,$K during the heating rate measurements. But even for stronger increase in $T_\text{f}$, the temperature scaling of the filter attenuation does not correlate with the heating rate data. Within the frequency range that is relevant for our experiment, $\omega_z=2\pi\times(0.4 - 1.8)\,$MHz, the low-pass filters show a slightly increasing attenuation for increasing temperature. This is likely due to an electric resonance caused by the parasitic inductance of the wiring and the low-pass filter capacitance. A noise source outside the cryostat penetrating through the low-pass filter lines would therefore produce a heating rate that decreases with rising temperature, in stark contrast to the behavior that we measure (Fig.\,\ref{fig:HR-vs-freq-below}).\\

\section{Ruling out Johnson noise for $T < \Tc$}
\label{app:Johnson-noise}
We exclude Johnson noise as dominant noise source for chip temperatures $T < \Tc$. First, we note that Johnson noise from the trap electrodes, bonding wires, and PCB traces, which are not filtered by the low-pass filters, should have a flat frequency dependence, see Eq.\,\eqref{eq:JN}. This is in clear contrast to the approximate $1/f$ scaling found in our data, Fig.\,\ref{fig:HR-vs-freq-below}. Second, we exclude Johnson noise from the low-pass filters using a temperature scaling argument. As the trap chip is heated to $T = \SI{100}{\kelvin}$, the filter temperature changes by only $\Delta T_{\text f} \approx \SI{2}{\kelvin}$ or roughly a factor \num{0.2}, due to the thermal insulation. Assuming a constant filter resistance in the range $\Delta T_{\text f}$, Johnson noise scales linearly with temperature, see Eq.\,\eqref{eq:JN}. The electric field noise produced by the filters should therefore increase by about a factor \num{0.2} as well. In contrast, the increase of the measured noise level in Tab.\,\ref{tab:TLF fit parameters} from $A=\SI{2.0+-0.1}{\volt^2\per\meter^2}$ at $T=\SI{12}{\kelvin}$ to $A=\SI{7.8+-0.3}{\volt^2\per\meter^2}$ at $T=\SI{77}{\kelvin}$ corresponds to a change by roughly a factor \num{3.9}, more than ten times larger than the change expected from Johnson noise from the filters.\\

In addition to the scaling arguments above, we calculate upper bounds for Johnson noise from trap electrodes, wiring, and low-pass filters. For a temperature $T=\SI{80}{\kelvin}$, each electrode is connected to a resistance $R_\text{tot}=R_\text{elec}+R_\text{wire}+R_\text{filter}\approx (102\text{ to }164)\,\si{\milli\ohm}$, where the individual contributions are calculated below. The electric field noise $S_E$ produced by the resistance $R_\text{tot}$ at the position of the ion is given by Eq.\,\eqref{eq:JN}. Using the individual electrodes' characteristic distances $\delta_\text{c}$ \cite{trap_sim} we arrive at a total level of expected field noise  
$S_{E,\text{80\,K}}^\text{(JN)} \approx \SI{6.0E-17}{\volt^2\meter^{-2}\hertz^{-1}}$ at $T=\SI{80}{\kelvin}$. We note that $S_{E,\text{80\,K}}^\text{(JN)}$ is an upper bound for the Johnson noise expectable at the three temperature sets in Fig.\,\ref{fig:HR-vs-freq-below}, since $R_\text{tot}$ will decrease at lower temperatures.
$S_{E,\text{80\,K}}^\text{(JN)}$ is roughly a factor 50 smaller than the noise corresponding to the smallest heating rate $\Gh\approx 0.3\,\text{phonons/s}$ we measure at $T=\SI{77}{\kelvin}$. Also, $S_{E,\text{80\,K}}^\text{(JN)}$ is still about an order of magnitude smaller than the smallest noise level $S_E = \SI{5.2 \pm 1.1 e-16}{\volt^2\meter^{-2}\hertz^{-1}}$ we measure at $T=\SI{12}{\kelvin}$. This shows that Johnson noise from these sources is negligible compared to the measured noise. The details of the calculation of $R_\text{tot}$ are given in the following. \\

Each of the trap's DC electrodes is connected to its first order RC filter via a gold wire bond connection and a gold-plated copper trace on the filter PCB. Electrodes C1, C2 are singly bonded, all other electrodes are doubly bonded. The wire bonds have a diameter of \SI{25}{\micro\meter} and a length of (1 to 2)\,\si{\centi\meter}. A single wire bond's resistance at $T=\SI{80}{\kelvin}$ is then $R_\text{wb} \approx \SI{50}{\milli\ohm}$, using a typical resistivity $\rho_\text{Au}\approx\SI{0.48E-8}{\ohm\meter}$ \cite{Mat79}. Typical values for contact resistances from chip to wire bond and from wire bond to PCB trace produced by our wedge bonder are $R_\text{wb-chip} \approx \SI{46.0+-0.2}{\milli\ohm}$, $R_\text{wb-PCB} \approx \SI{28.5+-0.2}{\milli\ohm}$, measured at room temperature in a 4-wire configuration. For the further calculation we assume that the contact resistances do not change with temperature. 

The traces have a width of \SI{300}{\micro\meter}, a thickness of \SI{100}{\micro\meter} and a maximal length of \SI{2}{\centi\meter} to the first filter capacitor. The trace thickness is larger than the skin depth in copper $\zeta_\text{Cu}\approx \SI{26}{\micro\meter}$ at $\omega=2\pi\times\SI{1.8}{\mega\hertz}$, calculated using Eq.\,\eqref{SM:skin-depth} with a typical resistivity $\rho_\text{Cu}\approx\SI{0.22E-8}{\ohm\meter}$  at $T=\SI{80}{\kelvin}$ \cite{Mat79} and $\mu=\mu_0$. Therefore we use twice the skin depth instead of the trace thickness to calculate the trace resistance $R_\text{tr} \approx \SI{3}{\milli\ohm}$ at $T=\SI{80}{\kelvin}$. The total resistance of the wiring connected to electrode C1 (or C2) at $T=\SI{80}{\kelvin}$ is then $R_\text{wire} = R_\text{tr}+R_\text{wb-PCB}+R_\text{wb-chip}+R_\text{wb}\approx \SI{126}{\milli\ohm}$. For all other electrodes the bond and contact resistances have to be replaced by half the value such that $R_\text{wire} \approx \SI{64}{\milli\ohm}$ because of the double bond connection.

The resistances $R_\text{f}$, $R_\text{a}$, $R_\text{b}$ within the RC filter circuit are another source of Johnson noise. The corresponding electric field noise is calculated by considering the effective real resistance $R_\text{filter} = R_\text{eff}$ of the circuit from the perspective of the trap electrode \cite{Bro15}. For the filter circuit shown in Fig.\,\ref{SM:filter-layout}\,(a) the effective real resistance is given by 
\begin{equation}
\label{SM:filter-impedance}
R_\text{eff}=\Re\left\lbrace\!\left(\frac{-i}{\omega C_\text{el}}\right)\!\parallel\!\left(R_\text{b}-\frac{i}{\omega C_\text{b}}\right)\!\parallel\!\left(R_\text{a}-\frac{i}{\omega C_\text{a}}\right)\!\parallel\! R_\text{f}\right\rbrace\,,
\end{equation}
where $a\parallel b$ denotes the impedance of two elements $a$, $b$ in parallel. 
The ESR of the filter capacitors is frequency dependent. Within the relevant frequency range $\omega_z=2\pi\times(0.4 - 1.8)\,$MHz the maximal ESRs are $R_\text{a}=\SI{24\pm1}{\milli\ohm}$ and $R_\text{b}=\SI{1.3\pm 0.1}{\ohm}$ according to the room temperature specification of the capacitors. This gives rise to a maximal effective real resistance $R_\text{filter} = R_\text{eff}=\SI{38\pm1}{\milli\ohm}$. 

We further give an upper bound for the amount of Johnson noise produced in the trap electrodes. In this calculation we neglect the influence of the electrodes' gold top layer, since the resistivity of gold is much higher than the resistivity of the YBCO film below it, which is small but finite in the RF domain, even below \Tc \cite{Supra_AC_losses}. The surface resistivity of the \SI{50}{\nano\meter} thick YBCO film at $f=\SI{10.9}{\giga\hertz}$ and $T=\SI{10}{\kelvin}$ is $\varrho_\text{YBCO}\approx\SI{0.66}{\milli\ohm}$ (specified value $\varrho_\text{YBCO}\approx\SI{0.1}{\milli\ohm}$ for \SI{330}{\nano\meter} thickness and $T=\SI{10}{\kelvin}$, $f=\SI{10.9}{\giga\hertz}$; Ceraco ceramic coating GmbH, Ismaning, Germany). Extrapolating the known quadratic scaling of the resistivity with frequency \cite{Mil91} down to the MHz regime, we calculate a surface resistivity $\varrho_\text{YBCO}=\SI{1.8E-11}{\ohm}$ at $f=\SI{1.8}{\mega\hertz}$ and $T=\SI{10}{\kelvin}$. Further assuming a temperature scaling $\varrho\propto (T/\Tc)^2/\sqrt{1-(T/\Tc)^4}$, \cite{Mik95}, we calculate a surface resistivity $\varrho_\text{YBCO}=\SI{3.4E-11}{\ohm}$ at $f=\SI{1.8}{\mega\hertz}$ and $T=\SI{80}{\kelvin}$. In comparison, the \SI{200}{\nano\meter} thick Au top layer even at $T=\SI{10}{\kelvin}$ still has a surface resistivity of $\varrho_\text{Au}=\SI{1.1}{\milli\ohm}$ \cite{Mat79}. From the YBCO surface resistivity we calculate the resistance of the trap electrodes for our trap geometry. We show here as an example the calculation for one of the meander-shaped electrodes, which have by far the largest resistance. These electrodes have a length $l=\SI{5.18}{\milli\meter}$ and a width $w=\SI{10}{\micro\meter}$. The total meander resistance at $f=\SI{1.8}{\mega\hertz}$ and $T=\SI{80}{\kelvin}$ is then $\Rm=l\varrho_\text{YBCO} /w=\SI{17.8}{\nano\ohm}$, which is 7 orders of magnitude smaller than the resistance $R_\text{wire}$ of the wiring. The resistances of the other trap electrodes are even smaller as an analog calculation shows. The electrodes' resistances $R_\text{elec}$ can hence be neglected.\\

\section{Influence of the YBCO meander electrodes on the ion heating rate for $T < \Tc$}
\label{app:meanders}
We exclude any other potential effects of the superconducting YBCO meanders connected to C1 and C2 on the ion heating rate below \Tc, like for instance electromagnetic pickup noise in the meander structure. For this we use a second, similar trap chip in which we compare the heating rate with electrodes C1 and C2 connected to the YBCO meanders (same configuration as for the experiment reported here) or directly attached to the low-pass filters. We find no difference between these two configurations, and observe in both cases a heating rate $\Gh=\SI{0.7\pm 0.1}{phonons\per\second}$ at $\omega_z = 2\pi\times\SI{1.0}{\mega\hertz}$ and $T=\SI{14}{\kelvin}$, comparable to the value $\Gh=\SI{0.23\pm 0.02}{phonons\per\second}$ at $\omega_z = 2\pi\times\SI{1.0}{\mega\hertz}$ and $T=\SI{12}{\kelvin}$ in Fig.\,\ref{fig:HR-vs-freq-below}.\\

\bibliography{BIB-PRA}
\end{document}